\documentclass[%
reprint,
amsmath,amssymb,
aps,
]{revtex4-1}

\usepackage{graphicx}
\usepackage{dcolumn}
\usepackage{bm}
\usepackage{hyperref}

\usepackage{color}

\usepackage{empheq}

\begin{document}
 
\title{Compressing phase space detects state changes in nonlinear dynamical systems}

\author{Valeria d'Andrea}
\email[Corresponding author:~]{vdandrea@fbk.eu}%
\affiliation{CoMuNe Lab, Fondazione Bruno Kessler, Via Sommarive 18, 38123 Povo (TN), Italy}
\author{Manlio De Domenico}
\email[Corresponding author:~]{mdedomenico@fbk.eu}%
\affiliation{CoMuNe Lab, Fondazione Bruno Kessler, Via Sommarive 18, 38123 Povo (TN), Italy}

\date{\today}

\begin{abstract}
Equations governing the nonlinear dynamics of complex systems are usually unknown and indirect methods are used to reconstruct their manifolds. In turn, they depend on embedding parameters requiring other methods and long temporal sequences to be accurate. In this paper, we show that an optimal reconstruction can be achieved by lossless compression of system's time course, providing a self-consistent analysis of its dynamics and a measure of its complexity, even for short sequences.  Our measure of complexity detects system's state changes such as weak synchronization phenomena, characterizing many systems, in one step, integrating results from Lyapunov and fractal analysis.
\end{abstract}

\maketitle

\section{\label{sec:sec1}Introduction}
Dynamics of natural systems is often described by nonlinear equations. When those equations are unknown, we can reproduce the system dynamics through the reconstruction of the manifold from the time course of one of its variables   \cite{Ott1994, Packard1980, Farmer1987}. Phase space reconstruction has been widely applied in modeling and predictions of several nonlinear systems, such as ecological, climate and neural ones  \cite{ Casdagli1989, Chavez2010, Sugihara2016a, Brunton2017}.
The embedding theory proposed by Takens \cite{Takens1981} allows one to reconstruct a one-to-one map of the attractor of a dynamical process using time-lagged values of a single system variable. The delay-coordinate map is built from the time series $X(t)$ by vectors in $R^m$ of the form $\textbf{X}_n = [X(n),X(n-\tau), x(n-2\tau),...,X(n-(m-1)\tau)] $, where $\tau$ is the time delay. To correctly build the embedding of $d$-dimensional manifold $\textbf{M}$ it is crucial to choose adequate values for $m$ and $\tau$, i.e. the embedding parameters.\\
According to the Whitney theorem, the diffeomorphism on $\textbf{M}$ is ensured by choosing an embedding dimension $m>2d+1$ \cite{Whitney1936} and the result may be  generalized also to non-integer (fractal) dimension \cite{Sauer1991}. Whitney theorem has been relaxed, for example in \cite{Haefliger1963,Fuquan1994}, but still those studies provide an upper bound for the estimation of $m$. 
Several methods were developed to estimate the minimum possible embedding dimension \cite{Cellucci2003} and usually those methods are based to the fact that, when evaluating some quantities on a $R^m$ delay-coordinate map, they do not vary for $m$ higher than the proper embedding dimension. Those diffeomorphism invariants could be, for examples, the largest Lyapunov exponent or the percentage of false nearest neighbours \cite{Kennel1992,Abarbanel1993}, where the latter option, in its implementation introduced by Cao \cite{Cao1997}, is currently the most used method to estimate the minimum $m$.\\ To estimate the embedding dimension, methods that involve the fact that entropies are diffeomorphism invariants have been proposed and include, for example, differential entropy \cite{Gautama2003} and permutation entropy \cite{Riedl2013}, where the latter has the advantage to take into account the temporal information contained in the time series \cite{Bandt2002}. Kolmogorov complexity, also known as algorithmic entropy, was proposed in 1968 as a measure of the amount of information of the trajectory of a dynamical process \cite{Kolmogorov1968} and is defined as the length of the shortest description that produces the trajectory as output. Even if Kolmogorov complexity cannot be computed, for the trajectories of a dynamical system it is usually approximated using lossless compression algorithms, following the theorem of Brudno who, in 1978, wrote the equality between Kolmogorov complexity and entropy rate \cite{Brudno1978}. Nevertheless, to date, estimating embedding dimension is still far from being an easy task, although this parameter is critical to gain insights about the physics of the underlying dynamical system.\\

In this paper we show that optimal embedding dimension can be estimated through a measure of the Kolmogorov complexity, that is here evaluated using the compression algorithm introduced by Lempel and Ziv \cite{Ziv1978}. Our dimension estimate could represent a more robust measure than other information estimators because is independent on the system representation \cite{Argenti2002} so it may be estimated without prior knowledge of the value of optimal time-delay $\tau$ \cite{Fraser1986}. The main advantage of our approach is that we explore the geometry of the manifold of the dynamical system with complexity measures that capture a rich information about the underlying dynamics and reveal change in the system state that are otherwise difficult to detect \cite{Melchert2015,Avinery2017,Martiniani2019}. In particular, here we show that exploring how the system approaches its proper embedding dimension can reveal the emergence of chaotic synchronization phenomena in a coupled drive-response system.\\

\section{\label{sec:sec2}Low-dimensional chaotic systems}
To estimate the optimal embedding dimension $m$, we  built $\textbf{M}_X (\tau, m)$, an ensemble of delay-coordinate maps from $X(t)$ as a function of time delay $\tau$ and $m$. Then, at fixed $\tau$ and $m$, we symbolized each degree of freedom so generating a new discrete variable with a scale that depends on the bin size $\epsilon$ used to make discrete the delay-coordinate map: $X_{discrete}=X_n(\epsilon, \tau,m)= (x_1, x_2,...,x_n)$. We computed the entropy rate of the resulting sequence of symbols through a Lempel-Ziv data compression algorithm \cite{Lin2012}:
$ S = (\frac{1}{n}\sum\limits_{i=2}^{n} \frac{L_i^i}{\log{i} })^{-1}$, where $L_i^n$ is the shortest sub-sequence starting at index $i$ that does not appear in the window $ x_{i-n}^{i-1}$ of length $n$. We evaluated entropy rate for the entire ensemble of delay-coordinate maps $\textbf{M}_X (m)$ and estimated as the optimal embedding dimension $m$ the one such that $\frac{S[\textbf{M}_X(m+1)]}{S[\textbf{M}_X(m)]}=cost $. 
This means that the optimum embedding dimension is the one at which entropy rate has at least a component that behaves as a non linear function of $m$, that is $S(m)\sim c_1 e^{c_2 m}$. That choice was suggested by the fact \cite{Bialek2001} that system with causal interactions among their elements have entropy that grows as a non extensive function of their size $S(N)=S_0N + S_1(N)$, where the non extensive component is described by a power law function $ S_1\sim N^m$.  \\

\begin{figure}
\includegraphics[width=\linewidth]{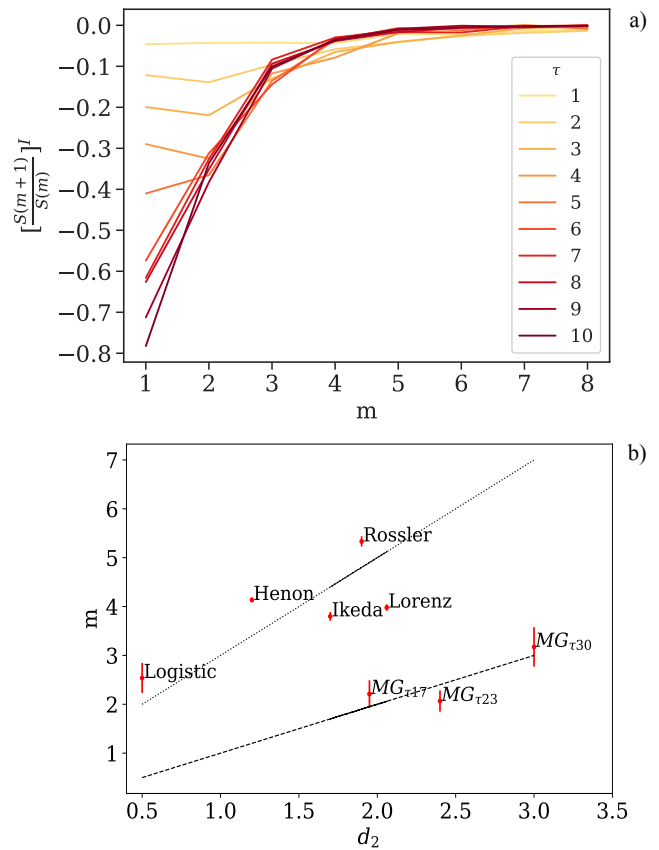}
\caption{Estimated embedding dimensions for low dimensional chaotic system. (a) First derivative of $\frac{S[\textbf{M}_X(m+1)]}{S[\textbf{M}_X(m)]}$ as a function of $m$ at different time delays $\tau$ for a Lorenz dynamical system. Across all $\tau$ values the derivative reaches zero for $m=4$. (b) estimated $m$ for a set of chaotic systems. For comparison purposes values are plotted as a function of correlation dimension $d_2$. For each dynamical system we show mean and standard error of the mean evaluated across $N=50$ realizations. Dashed line corresponds to $m=d_2$, dotted line is $m=2d_2+1$}
\label{fig:Fig1}
\end{figure}

To estimate the optimal dimension for the embedding, avoiding the evaluation of the optimal time delay $\tau$, we tested our algorithm with a specific set of $\tau$ values and found robust results with respect to the choice of this parameter. Fig.~\ref{fig:Fig1}a shows an example for a single realization of a Lorenz signal, where estimation of optimal $m$ does not change for different $\tau$ values.
Fig~\ref{fig:Fig1}b shows the results of our algorithm for a set of chaotic systems. Specifically, we consider Logistic, H\'enon and Ikeda maps, Rossler, Lorenz and Mackey-Glass systems with three different time delays, widely used to model the dynamics of several natural phenomena, from chemical reactions to climate. For each system we computed our measures across 50 different realizations and compare our estimates with correlation dimension ($d_2$) measures \cite{Grassberger1983_Jan, Grassberger_May, Grassberger1983_Sept, Grassberger1983_Oct}. We found that for most of the tested systems, our dimension estimate is close to the  Whitney's upper bound $2d_2+1$, while, for Mackey- Glass systems, that we tested at three different time delays, we found that our $m$ measures are close to the lower bound delimited by $d_2$.\\
  
\section{\label{sec:sec3}Undirectionally coupled systems}
In coupled chaotic systems with a drive-response configuration, generalized synchronization (GS) may occur if the state of response system $X$ does not depends on its initial condition but depends only on the state of the driver $Y$, that is, if there is a functional relation between trajectories in the phase-space: $X(t)=\Phi(Y(t))$. When $\Phi$ is the identity, there is identical synchronization, that is easy to detect because the synchronized motion becomes simply a sharp line in $X(t)$ vs $Y(t)$ plane \cite{Rulkov1995}. Otherwise, when $\Phi$ differs from the identity, weak GS may emerge and this phenomenon is difficult to detect. 
Different methods to detect GS have been proposed \cite{boccaletti2018}.

For instance, it has been proven that synchronization occurs when all of the conditional Lyapunov exponents are negative \cite{Pecora1990}, while it is possible to gain insight into the the kind of synchronization that is acting by considering the dimension of the global synchronization manifold $d_G$ with respect to the dimension of the driver system $d_D$: if $d_G=d_D$ then the response system does not have an effect on the global dimension and there is identical synchronization. Otherwise, if $d_G>d_D$, the global manifold has a fractal structure and the synchronization is weak \cite{Pyragas1997}. To reveal weak GS in a coupled system, two different classes of measures are needed, namely conditional Lyapunov exponents and dimension(s) of the global manifold.

Here we show that the analysis of the dimension of the response system through lossless complexity measures can easily detect the emergence of GS. To this aim, we studied synchronization phenomena between two unidirectional chaotic systems, where GS takes place as a function of coupling factor $C$. We studied the optimal dimension $m$ of the systems assuming, as we did for non coupled systems, that entropy is well described by a non extensive function of number of elements $S \sim N^m $. That assumption is especially well posed when the system is weakly sensitive to initial conditions, where it was proven \cite{Latora2000, Ponce2020} that he usual Shannon entropy measures are not appropriate, and a new measure of entropy has to be introduced, that depends on sensitivity to initial conditions and from the multifractal spectrum.

\subsection{\label{sec:sec3_1}Heterogeneous systems}
As a first example we considered an unidirectionally coupled system in which the autonomous driver $\textbf{X}$ is a Rossler oscillator:

\begin{empheq}[left = \empheqlbrace]{align}
    &\dot{x}_1=-6\{x_2 + x_3\} \nonumber \\
    &\dot{x}_2= 6\{x_1+0.2 x_2\}\\
    &\dot{x}_3= 6\{0.2+x_3(x_1-5.7)\}\nonumber
    \label{eq:one}
\end{empheq}

and the driven one, $\textbf{Y}$, is a Lorenz oscillator:

\begin{empheq}[left = \empheqlbrace]{align}
    &\dot{y}_1=10(-y_1+y_2) \nonumber \\
    &\dot{y}_2= 28 y_1 -y_2 -y1y_3+C x_2^2\\
    &\dot{y}_3= y_1 y_2-2.66y_3\nonumber
    \label{eq:two}
\end{empheq}

\begin{figure}
\includegraphics[width=\linewidth]{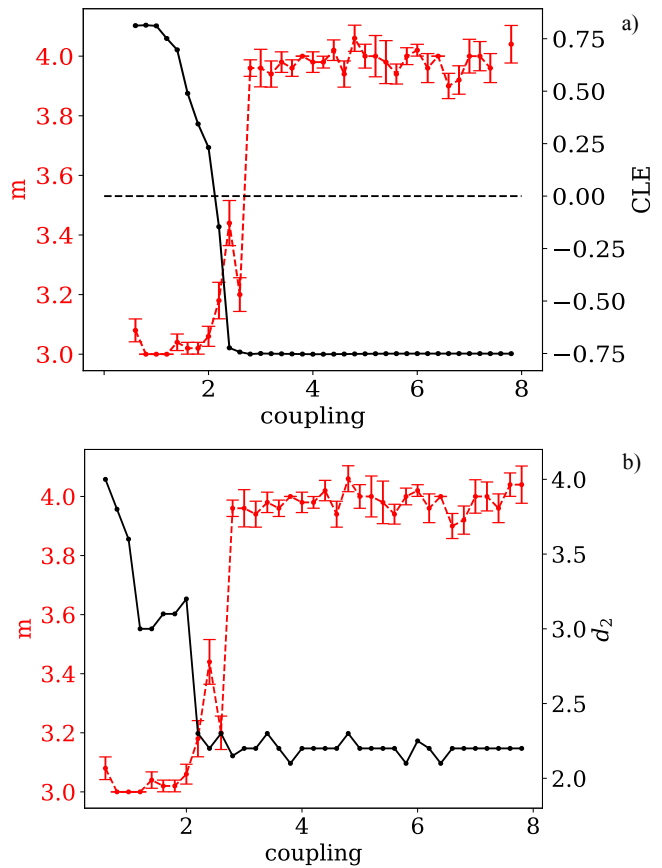}
\caption{ Lorenz driven by Rossler system. $(a)$ Estimated $m$ and correlation dimension $d_2$ as a function of coupling strength.  $(b)$ Estimated $m$ and conditional Lyapunov exponents}.
\label{fig:Fig2}
\end{figure}

 This type of system was investigated in previous works \cite{Pyragas1996,Quiroga2000,Quiroga2002}. In Fig.~\ref{fig:Fig2}a we show that, similarly to previous studies, GS arises for a threshold coupling strength $C=C_w>2.1$, where the conditional Lyapunov exponent becomes negative. We computed  Lyapunov exponents using the pull-back method \cite{Benettin1980,Wolf1985}, that relies on the Gram-Schmidt orthonormalization of Lyapunov vectors while integrating the dynamical system with  a fourth order Runge-Kutta algorithm (integration time step $dt = 0.01$). We computed exponents with 5000 time points, after discarding the first 10000 iterations. The correlation dimension $d_2$ is estimated by using 25000 time points and looking for the plateau in the function $d_2(m, \epsilon)$ \cite{Hegger1999}, indicating a suitable scaling relationship. As show in Fig.~\ref{fig:Fig2}b, $d_2$ of the global manifold is higher than $d_2$ of the driver Rossler system, indicating that at the  threshold $C_w$ the whole system undergoes a regime of weak synchronization.
 
 For each coupling value, we estimated the optimal embedding dimension as the average across 50 realizations of the system dynamics. Time series with 1000 time points were used for the estimation. As approaching the synchronization threshold $C_w$, $m$ increases abruptly and assumes values between the two extremes of two independent Lorenz and Rossler systems. Furthermore, is worth noting that the trend of $m$ estimates is opposite to the trends of both $d_2$ and conditional Lyapunov exponents, suggesting that those measures are referring to different but complementary properties of the dynamical system. Previous studies investigated how measures of entropy and complexity are both needed to describe natural systems, since they capture different properties of the dynamics \cite{Crutchfield2003,Feldman2008}. In particular, Lyapunov exponents and fractal dimension measures were usually related to the degree of randomness and disorder of the dynamics, while our hypothesis is that $m$, that is the dimension at which the entropy rate is described by a non linear function, is related to the length of the patterns, i.e., to regularities in the dynamics that allow for its compression.\\

\subsection{\label{sec:sec3_2}Identical systems}
 A second example we considered is the undirectionally coupled system formed by two identical H\'enon maps \cite{Quiroga2000}, where the driver is described by the system:
 
 \begin{empheq}[left = \empheqlbrace]{align}
    &\dot{x}_1=1.4-x_1^2 + 0.3 x_2 \\
    &\dot{x}_2= x_1 \nonumber
\end{empheq}

and the driven one by:

 \begin{empheq}[left = \empheqlbrace]{align}
    &\dot{y}_1=1.4-(C x_1 y_1 + (1-C)y_1^2) + 0.3y_2 \\
    &\dot{y}_2= y_1 \nonumber
\end{empheq}

\begin{figure}
\includegraphics[width=\linewidth]{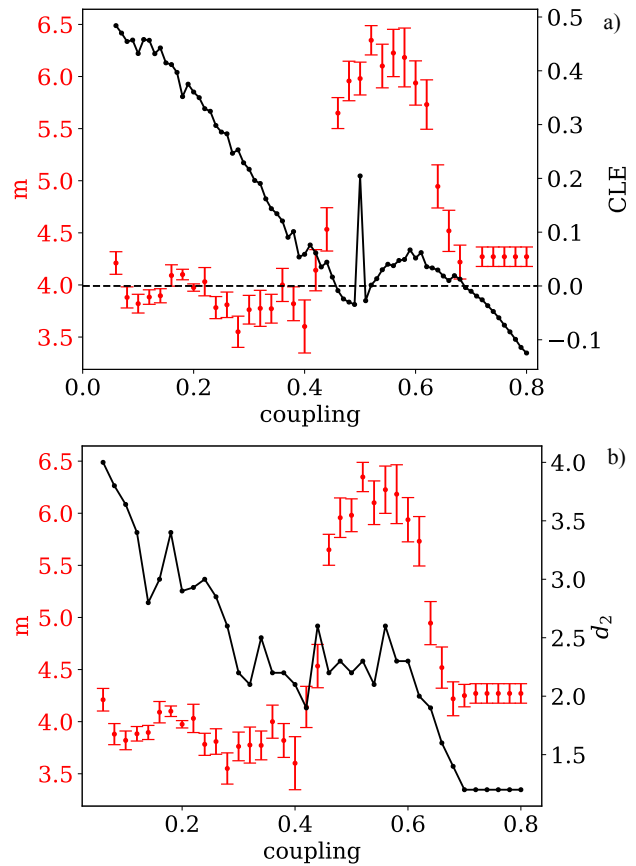}
\caption{ Coupled H\'enon maps. $(a)$, Estimated embedding $m$ and correlation dimension $d_2$ as a function of coupling strength.  $(b)$, Estimated $m$ and conditional Lyapunov exponents.}
\label{fig:Fig3}
\end{figure}

We computed Lyapunov exponents using the pull-back method with 5000 time points and we found that the conditional exponent takes negative values in two different intervals of couplings: in a window  $0.44<C_{w}<0.54$ and then for $ C_{i}>0.68$ (see Fig.~\ref{fig:Fig3}a). As shown in Fig.~\ref{fig:Fig3}b, in the first window $C_{w}$, the correlation dimension of the global manifold is higher than the correlation dimension of an independent H\'enon system: $d_G \approx 2.2 > d_{Henon}= 1.2$, indicating that the synchronization is weak in this interval. Furthermore, for coupling values higher than $ C_{i}$ we have that $d_G = d_{Henon}= 1.2$, showing that for high couplings identical synchronization takes place.
Both the coupling strength intervals and the two different regimes for GS are revealed with a single embedding measure. Here we computed for each coupling value the optimal embedding dimension $m$ as the average across 50 realizations, 1000 time points each, using lossless compression of the dynamics. We found that in $C_{w}$ interval the complexity of the coupled system increases, giving rise to an increase estimated $m$ of the global manifold. For $C> C_{i}$ the optimal $m$ has a drop, showing that there is a change in the system state, in particular $m$ estimates take values typical of an independent H\'enon map, revealing an identical synchronization.\\

\section{\label{sec:sec4}Conclusion}
In conclusions, we have shown that complexity measures used to reconstruct the geometry of the manifold of a dynamical system can be used to gain many insights about the system itself, even when the underlying governing equations are not known. We observed how the irregularity of the dynamics, expressed by entropy rate estimates, reaches a plateau and remains constant by increasing the dimension of the manifold, providing a robust and parameter-free estimate of the intrinsic optimal dimension. 
Our measure is quite stable for different values of time-delay $\tau$, providing a desirable method for the reconstruction of the  manifold that relies only on a single estimate.\\
We choose to relate complexity of the system to the way at which entropy rate measures departs from extensive functions and become non linear functions of the number of system dimensions. How to proper evaluate complexity has been a debated topic in last years. One of the most debated issue is the fact that information theoretic estimates like Shannon entropy measure the degree of randomness of the system and don't take into account system's dynamical organization, whereas ideal complexity measures should treat both random and lower distributions as minimally complex \cite{Deacon2014}. In our approach we focused on the entropy component that deviates from extensivity, arguing that it contains the information that has to be related to effective system's complexity\\
To detect synchronization usually are investigated quantities related to the randomness of the dynamics \cite{Abarbanel1996, Boccaletti2002}, such as Lyapunov exponents and fractal dimension. However, to be estimated in a reliable way, those quantities require long time series, in particular to compute correlation dimension, which is also potentially biased by user's choices about proper scale and dimensions. Our method, on the contrary, give robust results for shorter time series and has the advantage to capture and distinguish, with a single measure, different synchronization regimes.
Furthermore, the dimension at which the time series reaches its maximum disorder is informative and gives us insights about the intrinsic structure of the system. The way in which the optimal embedding dimension varies as a function of the parameters ruling the system dynamics highlights state changes, as long as they affect regularities in dynamical patterns. In this paper, we focused more specifically on the detection of generalized synchronization in coupled chaotic systems, a phenomena that appear in many biological and physiological processes \cite{Glass2001,Chen2017,Frasca2008}, as well as in geophysical fluid dynamics \cite{Duane2001}, but it is notoriously difficult to unravel. Additionally, the detection of synchronization phenomena permits the identification of causal drivers and leads to a better description and prediction of system dynamics. The key role that causal influence among observables has for the forecasting of their time course has been addressed in many studies related, for example, to ecological \cite{Ye2015}, financial \cite{May2008} and multi-scale human mobility systems \cite{Barbosa2018, DeDomenico2013}. Our method paves the way for applications to more complex dynamics exhibiting phenomena that usually require multiple complexity measures to be detected, showing that lossless compression of system's dynamics in the phase space can be suitably used for this purpose.\\


\bibliographystyle{apsrev4-2} 
\bibliography{biblio}

\end{document}